\documentclass[12pt]{article}

\usepackage{amsmath,amsthm,amssymb,color}
\usepackage{amsfonts,latexsym, hyperref, bbm, mathrsfs}
\begin{document}
	\title{{\bf{Back Reaction Inhomogeneities in Cosmological Parameter Evolution via Noncommutative  Fluid }}}
{\bf \maketitle}
	\vspace{0.0cm}
	\begin{center}
		Praloy Das \footnote{E-mail: praloydasdurgapur@gmail.com}and
		Subir Ghosh \footnote{E-mail: subirghosh20@gmail.com }\\
		\vspace{2.0 mm}
		\small{\emph{Physics and Applied Mathematics Unit, Indian Statistical
				Institute\\
				203 B. T. Road, Kolkata 700108, India}} \\
	\end{center}
	\vspace{0.5cm}
	
	\begin{abstract}
	The paper discusses back reaction effects in cosmology, (\'{a} la Buchert et. al.),  induced by noncommutative geometry effects in fluid. We have used generalizations of an action formulation of noncommutative fluid model, proposed earlier by us. We show that the noncommutative effects, depending on its nature, can play  either roles of (kinematical) dark matter or Cosmological Constant.
		
\end{abstract}		
		
	 \section{Introduction}
	 
	 Fluid dynamics is currently playing a central role in diverse aspects of modern physics such as fluid-gravity correspondence, cosmology, among others. This wide acceptance of fluid dynamics stems from the fact that it provides   a universal description of long wavelength physics that deals with low energy effective degrees of freedom of a field theory, classical or quantum and applies equally well at macroscopic and microscopic scales. Even though the two mathematical frameworks, {\it{i.e.}} discrete Lagrangian and continuum Hamiltonian (or Euler) formulations,  are extremely robust and have survived till date, in recent times  ways to tweaking these structures non-trivially at a fundamental level in a consistent way have been suggested. One such proposal put forward by us \cite{pra,das} is to introduce a Non-Commutative (NC) (or Non-Canonical, which might be more appropriate considering our classical scenario) phase space at the fundamental Lagrangian level and study how it modifies the field theoretic Euler fluid dynamics. (Fluid dynamics from the perspective of High Energy Physicists has been discussed in detail in the lucid review \cite{jac}{\footnote{Some subtle aspects of Eulerian fluid dynamics have appeared recently in \cite{arpan}}}.) We have termed our generalized system as NC fluid that consists of an extended fluid variable algebra along with a Hamiltonian that generates a modified form of fluid dynamics, the continuity and Euler (or force) equation. The new system involves a constant antisymmetric NC parameter $\theta_{ij}=-\theta_{ji}$ and reduces smoothly to the conventional fluid model for $\theta_{ij}\rightarrow 0$. Several novel and interesting properties of the NC model were revealed in \cite{das} (see also \cite{ar1}). In the present paper we have extended the previous work \cite{das} to include fluid vorticity but more importantly we have studied implications of the NC fluid model in cosmology. In particular we show that NC contributions can induce inhomogeneity in standard 
Friedmann-Robertson-Walker (FRW) cosmology that can impact the vital theme of structure formation.

Let us spend a few words on the other (comparatively) recent development in quantum and classical physics: introduction of extended forms of phase space Heisenberg commutators or its classical counterpart, non-canonical extensions of Poisson  brackets. Generically these new structures are referred to as NC algebra. NC generalizations of conventional theories became very popular after the work of Seiberg and Witten \cite{sw} who showed that in certain low energy regime String theory mimics an NC version of quantum field theory. Subsequently there have appeared numerous studies regarding effects of NC extensions in quantum and classical physics (for reviews see \cite{ncrev}). 

Broadly there are two inequivalent ways to incorporate NC effects in a theory. In one formulation in a field theory action the products of fields are replaced by $*$ (star) products that introduces NC contributions perturbatively in the action. This approach has been utilized in some previous works of NC fluid models \cite{van} where $*$-products are used in directly Euler fluid Hamiltonian (or equivalently in the action that generates Eulerian dynamics). 

We, on the other hand, follow an another route which (at least to our mind) is more direct with clear interpretations. We exploit the conventional map between Lagrangian (discrete fluid particle coordinates)   degrees of freedom and Euler (continuous field theoretic) variables consisting of density and velocity fields \cite{jac}. We introduce NC in Lagrangian coordinates and rederive the extended Euler algebra that involves NC effects. Finally we use this NC Eulerian structure to obtain fluid equations of motion in a Hamiltonian framework \cite{pra, das}. Another variation of our formalism has been pursued in \cite{Kai} with results a little different from ours.

The main focus of the present work is the following. We convert the NC fluid equations to co-moving frame and utilizing conventional techniques generate the NC extension of FRW-like equations in Newtonian cosmology. The results clearly show that NC effects can lead to inhomogeneities and anisotropy. Here we encounter a serious technical issue, that of spatial averaging in the context of cosmology. Buchert and Ehlers \cite{buc,buc1} and Buchert \cite{buc2} have developed a rigorous technique of averaging in cosmology and have steered the interest towards  analysis of fluctuations (see also \cite{buc4}) that can have significant impact "locally" that is in a domain small compared to horizon (whereas global averages will generically reduce to surface terms that vanish for boundaryless (three torus) Newtonian cosmology). We have followed the above formalism and have computed explicit forms for the NC-induced fluctuations.
	 	 
Finally we come to the most important part, i,e, the perspective and motivation of the present work. 	Cosmology is confronted with the following deep question: is  the apparent homogeneity and isotropy of  matter in the universe as well as the  observed isotropy of the Cosmic Microwave Background (CMB) enough to guarantee the Cosmological Principle which in turn ensures the validity of Friedmann-Robertson-Walker (FRW) framework. Obviously FRW approach makes the problem tractable since it involves only a single scalar dynamical variable, the scale factor $a(t)$, that  depends only on time.  Clearly  the presence of 
shear and vorticity can lead to  anisotropy via the celebrated the Sachs-Wolfe effect. But more interestingly and relevant to our context, is the possibility that a shear-less homogeneous background may also exhibit anisotropic curvature via the introduction of a canonical, massless 2-form field \cite{anis,12,anis1,anis2}. Furthermore it has been  shown \cite{19} that a canonical massless Kalb-Ramond field is a viable candidate for such a field. This class of  spacetimes are endowed with a   preferred direction in the sky along with a CMB that is isotropic at the background level.
Hence the anisotropy emerges at the level of the curvature of the homogeneous spatial sections, whereas the expansion is dictated by a single scale factor. Such partially homogeneous but anisotropic solutions of Einstein
equations having an isotropic expansion (describable  by a single scale factor). The anisotropy of these solutions are induced by 
the spatial curvature of the sections of constant comoving time where the curvature  is direction dependent. For a three-dimensional manifold this can be performed by considering spatial sections which are 
Cartesian product of curved subspaces that are generically taken as  homogeneous locally rotationally symmetric
class of metrics ({\it {e.g.}} Bianchi type III (B III) and Kantowski-Sachs (KS) solutions).
The above mentioned imperfect fluid generates an anisotropic stress that yields  a shear-free anisotropic model where the  scale
factor  evolves exactly like that  in a curved FRW
model. Subsequently  both redshift and the Hubble
parameter become. Even with an isotropic comoving distance
it was shown that  both the angular
diameter and luminosity distances reveal  anisotropy. Hence the anisotropy effects generated by  curvature can possibly be  detected   by such distance relations using, say, supernovae (SNe) \cite{19,nvvmb1}.

In this background let us finally discuss possible observational consequences of the results presented in our model. To compare and contrast with the above mentioned scenario we note that the tensorial constant NC parameter $\theta_{ij}$ can be identified with the two-form field introduced above but instead of exploiting the locally rotationally symmetric
class of metrics as above we have considered NC-extended FRW metric. As we discuss later in the paper, there are essentially two aspects where noncommutativity in the fluid can have effects, NC-corrected effective curvature and NC-corrected effective energy budget. Furthermore,   NC-effect can contribute additionally as either  Dark Matter or Dark Energy.

	  The paper is organized as follows: In section 2 we have computed the NC fluid algebra and the equations of motion. This is an extension of our earlier work \cite{das} since now we consider fluid vorticity as well. Section 3 deals with the construction of NC FRW equations. Section 4 is devoted to the study of averaging hypothesis in cosmology and derivation of NC induced fluctuation terms. Section 5 consists of noncommutative corrections to cosmological parameters. We conclude and suggest future directions of work in Section 6. An Appendix contains details of Dirac Bracket computations.

	 \section{Noncommutative Fluid Algebra and Darboux Map}
	 There are two ways of introducing noncommutativity in continuum fluid model: \\
	 (a) Treat the fluid field theoretic Hamiltoninan $H=\int d^3x~(\frac{1}{2}\rho v^2 +U(\rho))$ as the starting point, where $\rho(x),v_i(x)$ are density and velocity fields respectively and $U(\rho)$ is the barotropic potential that depends only on the density, in Eulerian framework. Now replace the local product of  fields, {\it {e.g.}} by star ($*$) product {\it {e.g.}} $\rho v_iv_i\rightarrow \rho~*v_i~*v_i $ where a generic $*$-product is formally  defined as $A(x)*B(x)=A(x)~exp(i\frac{1}{2}\theta_{ij}\partial_i^\leftarrow \partial_j^\rightarrow )~B(x)$, where $\theta_{ij}$ is the NC parameter. Upon expanding the exponential operator in powers of $\theta_{ij}$ the NC contribution will appear as additional higher (spatial) derivative terms in the Hamiltonian. Incidentally the NC algebra follows from the $*$-commutator $[x_i,x_j]_*=x_i*x_j-x_j*x_i=i\theta_{ij}$. From the above NC-extended Hamiltonian one can generate NC-modified continuity and Euler equation for NC fluid in the conventional way. This approach has been adapted in \cite{van}.\\
	 (b) We, on the other hand, follow a new approach, initiated by for the first time in \cite{pra,das,ar1} exploiting ideas provided in \cite{jac}. Indeed, this is possible due to the unique feature of fluid where, on one hand in the Lagrangian framework it can be interpreted as a collection of discrete particles whereas on the other hand in the Hamiltonian (or Eulerian) framework it is considered as a continuum field theory comprising of $\rho(x),v^i(x)$, the Eulerian field degrees of freedom. There is an explicit map that connects the Lagrangian (discrete variables, particle coordinate and velocity $X_i(x),\dot X_i(x)$ in the continuum limit) and Hamiltonian (continuous variables, density and velocity $\rho(x),v^i(x))$) degrees of freedom given by	
	 \begin{eqnarray}
	 \rho({\bf r})=\rho_0\int\delta(X(x)-r)dx,\label{nc1} \\
	 v^i({\bf r})=\frac{\int dx \dot{X_i}(x)\delta (X(x)-r)}{\int dx \delta(X(x)-r)}.
	 \label{nc2}
	 \end{eqnarray}	
	  Generalizing the canonical Poisson brackets between discrete variables
	 	  \begin{equation}
	  [\dot{X}^i , X^j ] = (i/m) \delta^{ij},~~ [X^i , X^j ] = 0,~~ [\dot{X}^i , \dot{X}^j ] = 0.
	  \label{nnp1}
	  \end{equation}
	  to the Lagrangian fluid variables,
	  \begin{equation}
	  \{X_i({\bf x}),\dot X_j({\bf x'})\}=\frac{1}{\rho_0}\delta_{ij}\delta({\bf x} - {\bf x'}),~ \{X_i({\bf x}), X_j({\bf x'})\}= \{\dot X_i({\bf x}),\dot X_j({\bf x'})\}=0,
	  	  \label{pb}
	  \end{equation}
	  with $\rho_0$ a constant background density,
	   it is straightforward to derive the canonical (Hamiltonian) algebra between the fluid variables
	   \begin{eqnarray}
	   \{\rho ({\bf x}), \rho ({\bf x}')\}=0,~ 
	   \{v^i({\bf x}), \rho({\bf x'})\} = \partial_i \delta (\bf x - \bf x'), \qquad \label{rrv}\\
	   \{v^i({\bf x}), v^j({\bf x'})\} = -\frac{\omega^{ij} ({\bf x})}{\rho({\bf r})} \delta (\bf x - \bf x'),
	   \label{rv}
	   \end{eqnarray}
	   where
	   \begin{equation}
	   \omega^{ij} ({\bf x}) = \partial_i \ v^j {(\bf x)}-\partial_j v^i ({\bf x})
	   \label{v}
	   \end{equation}
	   is called the fluid vorticity.

	    We find this approach more appealing since it is natural to introduce NC-algebra at the discrete coordinate level via 
	    	    \begin{equation}
	    [\dot{X}^i , X^j ] = (i/m) \delta^{ij},~~ [X^i , X^j ] = i\theta ^{ij},~~ [\dot{X}^i , \dot{X}^j ] = 0.
	    \label{nnp}
	    \end{equation}
	    with the generelization to NC Lagrangian fluid variable algebra,
	    \begin{equation}
	    \{\dot{X}^i ({\bf x}), X^j ({\bf x'})\} = \frac{1}{\rho_0} \delta^{ij} \delta ({\bf x} -{\bf x'}),~~ \{X^i ({\bf x}), X^j ({\bf x'})\} =\frac{1}{\rho_0} \theta ^{ij}\delta ({\bf x} -{\bf x'}),~~ \{\dot{X}^i ({\bf x}), \dot{X}^j ({\bf x'})\} = 0.
	    \label{p}
	    \end{equation}
	    In an identical fashion this yields NC fluid algebra,
	    	    \begin{eqnarray}
	    \{ \rho ({\bf x}),\rho ({\bf  x'})  \}=-\theta^{ij}\partial_i\rho({\bf x})\partial_j \delta ({\bf x} -{\bf x'}) ,
	    \label{e2}
	    \end{eqnarray}
	    \begin{eqnarray}
	    \{ v^i({\bf x}),\rho ({\bf  x'})\}=\partial _i \delta ({\bf x} -{\bf x'}) -\theta^{jk}\partial _j v^i({\bf x}) \partial_k \delta ({\bf x} -{\bf x'}) ,
	    \label{e1}
	    \end{eqnarray}
	    \begin{eqnarray}
	    ~~~~~~~~~~~~ \{v^i({\bf x}),v^j({\bf x'})\}&=&\frac{(\partial_jv^i-\partial_iv^j)}{\rho}\delta ({\bf x} -{\bf x'}) \nonumber \\
	    &&+\theta^{kl}[\partial_l\delta ({\bf  x}-{\bf x'}) (\frac{\partial_k(v^iv^j)}{\rho}-v^iv^j\partial_k(\frac{1}{\rho}))\delta ({\bf x} -{\bf x'})+.....
	    \label{ee2}
	    \end{eqnarray}
	   	 But in \cite{das} we have made further progress by proposing an action that can generate the NC fluid algebra through the identification of Dirac brackets obtained from the NC action {\footnote{In this context we note that similar to current algebra in other models such as the Schwinger condition in a generic relativistic quantum field theory \cite{schwin}, anomalous chiral current algebra in fermionic models \cite{jackiw} among others, the bracket between  $0$'th component has a special status since it can be reproduced uniquely using different schemes. On the other hand rest of the current algebra are in general scheme dependent and does not enjoy such importance. The action posited by us induces correctly the zeroth component charge density algebra but does not completely reproduce rest of the algebra.}}.

	 We start with a Lin \cite{lin} (see also \cite{eck}) form of first order fluid action endowed by NC correction terms,
	 \begin{eqnarray}
	 L=-\partial_t\theta (\rho -\frac{1}{2}\theta^{ij}\partial_i\rho\partial_j\theta ) -(\frac{1}{2}\rho v^2 +U(\rho) )- \rho\alpha\partial_t\beta ,
	 \label{a1}
	 \end{eqnarray} 
	 where $v^i=\partial_i\theta+\alpha\partial_i\beta$ is the velocity field in Clebsch parameterization \cite{cl}. (For an authoritative monograph on fluid dynamics see \cite{lamb}. Note that in \cite{das} we had $v^i=\partial_i\theta(x)$, which amounts to the no vorticity condition. Conventional fluid dynamics is recovered for $\theta^{ij}=0$.

		 Let us first derive the equations of motion by varying  
		 $\rho$ and $v$ in the action:
		 \begin{eqnarray}
		 	\partial_t{\rho}=-\partial_i\left((\rho v^i)+\theta^{ij}[\rho\partial_j(\frac{v^2}{2})+\frac{1}{2}\partial_j\theta\partial_k(\rho v^k)-\rho\partial_j(\alpha v^k\partial_k\beta)+ \rho\partial_j U^\prime]\right)
		 	\label{b1}
		 \end{eqnarray}
		 \begin{eqnarray}
		 \partial_t{v^l}&=&-\partial_l(\frac{v^2}{2}+U^\prime)+\theta^{ij}\left[\frac{1}{2}\partial_l\partial_j\theta \partial_i(\frac{v^2}{2}+U^\prime)+\frac{1}{2}\partial_j\theta \partial_i(\partial_l(\frac{v^2}{2}+U^\prime))-\frac{1}{2}\partial_l\partial_j\theta \partial_i(\alpha v^k \partial_k\beta)\right] \nonumber \\
		 &&+\theta^{ij}\left[-\frac{1}{2}\partial_j\theta \partial_i(\partial_l(\alpha v^k \partial_k\beta))+\frac{\alpha}{\rho}\partial_i\rho\partial_l\beta\partial_j(\frac{v^2}{2}+U^\prime)\right] \nonumber \\
		 &&+\theta^{ij}\left[\frac{1}{2}\frac{\alpha}{\rho}\partial_j\theta\partial_i(\partial_k(\rho v^k \partial_l\beta))-\partial_i\rho\partial_j(\frac{\alpha}{\rho})\alpha v^k \partial_l\beta\partial_k\beta-\frac{\alpha^2}{\rho^2}\partial_i\rho\partial_j(\rho v^k \partial_l\beta\partial_k\beta) \right].
		 \label{z1}
		 \end{eqnarray}
		 A few comments are in order.\\
		 (i) This is a non-trivial extension to our earlier work \cite{das} as $\alpha, \beta$ variables are included indicating that the canonical fluid possesses a vorticity.\\
		 (ii) Notice the $U'$-dependent term in (\ref{b1}). Its contribution vanishes due to antisymmetry of $\theta_{ij}$ which we have followed in earlier works \cite{pra,das} and are presently adhering to. However, in an alternative formulation suggested in \cite{Kai} it can yield a non-zero contribution.\\

		  Let us exploit the Dirac bracket formalism \cite{dirac} to obtain,  to first non-trivial order in $\theta^{ij}$, the NC fluid algebra. (The detailed constraint analysis is provided in the Appendix ) The complete NC fluid algebra is given by,
		\begin{eqnarray}
					\{\rho (x),\rho(y)\}=-\theta^{ij}\partial_i\rho (x)\partial_ j^x \delta(x-y), ~~
			\{\rho (x),\theta(y)\}=\delta(x-y)+\frac{1}{2}\theta^{ij}\partial_j\theta(x)\partial_i^x\delta(x-y), ~~\nonumber \\
			\{\alpha(x),\rho(y)\}=\theta^{ij}\frac{\alpha(x)}{\rho(x)}\partial_i\rho(x)\partial_j\delta(x-y),~~
			\{\alpha(x),\theta(y)\}=-\frac{\alpha(x)}{\rho(x)}\delta(x-y)-\frac{\theta^{ij}}{2}\frac{\alpha(x)}{\rho(x)}\partial_j\theta(x)\partial_i\delta(x-y),~ \nonumber \\
			\{\alpha(x),\alpha(y)\}=-\theta^{ij}\frac{\alpha(x) }{\rho(x)}\partial_i\rho(x)\left(\frac{\alpha}{\rho}\partial_j\delta(x-y)+\partial_j(\frac{\alpha}{\rho})\delta(x-y)\right),~~~~ \nonumber \\
			\{\alpha(x),\beta(y)\}=\frac{\delta(x-y)}{\rho(x)},~~\{\rho(x),\beta(y)\}=\{\theta(x),\beta(y)\}=	\{\theta (x),\theta(y)\}= \{\beta(x),\beta(y)\}=0.
			\label{a2}
		\end{eqnarray}

		A non-vanishing charge density $\{\rho (x),\rho(y)\}$ bracket of a similar structure first appeared in \cite{jacnc,jac} which, however, was derived in a somewhat heuristic way in a completely different system, lowest Landau level in magnetohydrodynamics. The complete NC bracket structure provided here is new. Quite obviously this algebra  is more involved that our earlier results \cite{das} due to the presence of $\alpha, \beta $.
		
		We consider the conventional form of Eulerian fluid Hamiltonian,
			\begin{eqnarray}
			H=\frac{1}{2}\rho v^2 +U(\rho).
			\label{ham}
			\end{eqnarray}
		It is straightforward to check that, using the above Dirac bracket algebra (\ref{a2}), the Hamiltonian equations of motion,
		 	\begin{eqnarray}
	\dot{\rho}=\{\rho, H\},~~
	\dot{v^l}=\{v^l,H\} ,
	\label{z3}
	\end{eqnarray}
	agree with the previously computed dynamical 	equations (\ref{b1},\ref{z1}) obtained from action principle. This cross checking ensures overall consistency of the procedure. 		 These modified equations of motion constitute our  primary major results and the starting point of the present analysis.	\\
		
	In a Hamiltonian structure there is a very well known and convenient transformation known as the Darboux transformation \cite{dar} that allows (at least locally) a construction of the non-canonical variables in terms of canonical variables. In our case, operationally this simply means that  using Darboux map, the NC variables $\rho, \theta, \alpha, \beta$ can be expressed (at least locally) in terms of a canonical set $\rho_c, \theta_c, \alpha_c, \beta_c $ obeying canonical algebra, 
	$$\{{\rho_c} (x),{\rho_c}(y)\}=\{{\theta_c} (x),{\theta_c}(y)\}=\{{\alpha_c} (x),{\alpha_c}(y)\}=\{{\beta_c} (x),{\beta_c}(y)\}=0$$
\begin{eqnarray}
\{{\rho_c} (x),\theta_c(y)\}=\delta(x-y),~~\{{\alpha_c} (x),\theta_c(y)\}=-\frac{\alpha_c}{\theta_c}\delta(x-y),~~\{{\alpha_c} (x),\beta_c(y)\}=\frac{\delta}{\rho_c}.
\label{b2}
\end{eqnarray}. 
The explicit form of Darboux map to $O(\theta)$,  is given by 
\begin{eqnarray}
{\rho}=\rho_c+\frac{1}{2}\theta^{ij}\partial_i\rho_c\partial_j\theta_c;~~{\theta}=\theta _c ;~~{\beta}=\beta _c;~~{\alpha}= \alpha_c-\frac{\theta^{ij}}{2}\alpha \frac{\partial_i\rho\partial_j\theta}{\rho}
\label{b3}
\end{eqnarray} 
such that the NC algebra in (\ref{a2}) is reproduced. For simplicity we will just keep the notation $\rho,\theta,\alpha,\beta$ instead of $\rho_c, \theta_c, \alpha_c, \beta_c $. The  Hamiltonian (\ref{ham}) is now written in terms of canonical variables, (to order of $\theta^{ij}$),
 \begin{eqnarray}
 H= 
 \int dr [T-\frac{1}{2}\theta^{ij}\frac{\partial_j\rho \partial_i\theta}{\rho}(\frac{1}{2}\rho(\partial_i\theta)^2-\frac{1}{2}\alpha^2(\partial_i\beta)^2+U+P_c)]
 \label{e3}
 \end{eqnarray}
 where,  $v^i=\partial_i\theta+\alpha\partial_i\beta $, $T=\frac{1}{2}\rho v^2 +U(\rho)$ is the canonical energy density  and $P=\rho U^\prime-U$ is the   pressure. The continuity equation is obtained as,
 \begin{eqnarray}
 \dot{\rho}&=& \{\rho,H\} \nonumber \\
 &=&-\partial_l[\rho(v^l-\frac{1}{2}\theta^{lj}\partial_j\rho\frac{1}{\rho^2}(\frac{1}{2}\rho (\partial_i\theta)^2-\frac{1}{2}\alpha^2(\partial_i\beta)^2+U+P_c)-\frac{1}{2}\theta^{ij}\frac{(\partial_j\rho)}{\rho}\partial_i\theta\partial_l\theta )] \nonumber \\
 \label{e4}
 \end{eqnarray}
 is written in a suggestive form $\dot{\rho}=-\partial_l(\rho \bar {v}^l)$ where, \\
  $
 \bar {v}^l=(v^l-\frac{1}{2}\theta^{lj}\partial_j\rho\frac{1}{\rho^2}(\frac{1}{2}\rho (\partial_i\theta)^2-\frac{1}{2}\alpha^2(\partial_i\beta)^2+U+P_c)-\frac{1}{2}\theta^{ij}\frac{(\partial_j\rho)}{\rho}\partial_i\theta\partial_l\theta ) ,$
 is identified as the NC corrected generalized velocity.
The Euler equation can also computed in a similar way.	

Several features of the NC fluid system need to be stressed:\\
i) The NC Hamiltonian and NC Eulerian equations are expressed entirely in terms of {\it{canonical}} fluid variables.\\
ii) There exists a modified local conservation law of matter.\\
iii) In our formulation since the matter density $\rho $ is unchanged the total mass is same as in conventional case although the momentum flux receives NC corrections.
	
	 \section{Modification in Friedmann Equation}
Let us now move on to our present area of interest, NC modified cosmology.	The "Standard model" of cosmology consists of the continuity  equation and 
 Friedmann equation, (without Cosmological Constant),
\begin{eqnarray}
\dot{\rho}=-3H(\rho+P)=-3\frac{\dot{a}}{a}(\rho+P),
\label{nc14}
\end{eqnarray}
\begin{eqnarray}
\frac{\ddot{a}}{a}=-\frac{\rho+3P}{6M^2}.
\label{nc15}
\end{eqnarray}
$P$ denotes the pressure and $a(t)$ is the scale factor. $M=(8\pi G)^{-1/2}$ refers to Newton's constant $G$. Introducing the Hubble parameter $H(t)={\dot a}/a$ an equivalent equation follows:
\begin{eqnarray}
\frac{\dot{a}^2}{a^2}=H^2=\frac{\rho}{3M^2}-\frac{k}{a^2},
\label{nc16}
\end{eqnarray}
with a scaled $k=0,\pm 1$ indicating flat, closed or open universe respectively. Clearly inhomogeneity or anisotropy are not supported (see for example  \cite{lid}).

It is well known that one can rewrite the conventional fluid dynamical equations in comoving frame such that they agree with the FRW equations. This will be our starting point. Since we have developed a NC-extended set of generalized fluid equations its mapping to comoving coordinates will give rise to NC extended FRW dynamics. Since from now on in this paper we are primarily interested in cosmological scenario we will consider fluid without vorticity as canonical vorticity does not play any major role in cosmology.

The comoving coordinates are defined as
\begin{equation}
{\bf r} =a(t){\bf x}(t),
\label{c1}
\end{equation}
where ${\bf r}(t), {\bf x}(t)$ and $a(t)$ denote the proper coordinates, comoving coordinates and the  scale factor respectively. The laboratory velocity $\bf{v}$ is written as,
\begin{equation}
\dot{\bf r}= H(t){\bf r}+a\dot {\bf x}(t) ~ \rightarrow 
{\bf v}= H(t){\bf r} +{\bf u},
\label{cc2}
\end{equation}
with ${\bf u}$ defined as the peculiar velocity. In standard FRW ${\bf u}$ is taken as zero and it is considered as a perturbation.

We start from the simplified form of NC continuity and Euler equation, (\ref{b1}) and (\ref{z1}), with vorticity free  $v^i= \partial_i\theta$ only (ignoring the  $\alpha$ and $\beta$): 
\begin{eqnarray}
\frac{\partial \rho}{\partial t}|_r +\frac{\partial}{\partial r_i}\left(\rho v^i+\frac{1}{2}\theta^{ij}\rho \partial_j v^2+\frac{1}{2}\theta^{ij}\partial_k(\rho v^k)\partial_j\theta+\theta^{ij}\rho \partial_j U^\prime\right)=0,
\label{y1}
\end{eqnarray}
\begin{eqnarray}
\frac{\partial v_i}{\partial t}|_r+v_j\partial_j v_i=-\frac{\partial _i P}{\rho}-\partial_i \varPhi+\theta^{mj}\left[\frac{1}{2}\partial_i\partial_j\theta \partial_m(\frac{v^2}{2}+U^\prime)+\frac{1}{2}\partial_j\theta \partial_m(\partial_i(\frac{v^2}{2}+U^\prime))\right].
\label{c2}	
\end{eqnarray} 
We have introduced $\varPhi$ as a generic potential. We need to recast the dynamics in the comoving coordinates ${\bf {x}},t$. The space derivatives are easily related by
$$\partial /\partial {\bf{r}} =(1/a)\partial /\partial {\bf{x}}.$$ On the other hand, the time derivatives at constant ${\bf {r}}$ and constant  ${\bf {x}}$ are related by,
$$\frac{\partial}{\partial t}\mid_{\bf{r}}=\frac{\partial}{\partial t}\mid_{\bf{x}}-\frac{\dot a}{a}({\bf{x}}.{\partial_{\bf{x}}}).$$ 
Using the above identities we derive the cherished expressions for the NC fluid dynamics in comoving frames:
\begin{eqnarray}
\frac{\partial \rho}{\partial t}|_x+\frac{1}{a}\frac{\partial}{\partial x_i}(\rho u_i)+\frac{3\rho \dot{a}}{a}+\frac{\theta^{ij}}{a^2}\frac{\partial\rho}{\partial x_i}(\dot{a}^2x_j+\dot{a}u_j+(\dot{a}x_k+u_k)\frac{\partial u_k}{\partial x_j}) \nonumber \\
+\theta^{ij}\frac{1}{2a^3}\frac{\partial}{\partial x_i}(\frac{\partial}{\partial x_k}(\rho \dot{a} x_k+u_k\rho))\frac{\partial \theta}{\partial x_j}+\theta^{ij}\frac{1}{a^2}\frac{\partial\rho}{\partial x_i}\frac{\partial U^\prime}{\partial x_j} = 0,
\label{c6}
\end{eqnarray}
\begin{eqnarray}
\frac{\partial u_i}{\partial t}+\frac{\dot{a}}{a}u_i+\frac{u_j}{a}\frac{\partial u_i}{\partial x_j}+\frac{1}{a\rho}\frac{\partial P}{\partial x_i}+\frac{1}{a}\frac{\partial \phi_{pec}}{\partial x_i} +(\ddot{a}+\frac{4\pi}{3}aG\rho)x_i \nonumber \\
-\frac{\theta^{mj}}{a^3}\left[\frac{1}{2}\partial_i\partial_j\theta \partial_m((\dot{a} x_l+u_l)^2+U^\prime)+\frac{1}{2}\partial_j\theta \partial_m(\partial_i((\dot{a} x_l+u_l)^2+U^\prime))\right] = 0 .
\label{c7}
\end{eqnarray}
We can make  a further simplification by dropping the  peculiar velocity ${\bf{u}}$-dependent terms thus reducing the NC continuity equation (\ref{c6}) to 
\begin{eqnarray}
\dot{\rho}+\frac{3\rho\dot{a}}{a}+\Psi=0,~~\Psi=\theta^{ij}[\frac{1}{a^2}\frac{\partial \rho}{\partial x_i}\dot{a}^2 x_j+\frac{1}{2a^3}\frac{\partial}{\partial x_i}(\frac{\partial}{\partial x_k}(\rho \dot{a} x_k))\frac{\partial \theta}{\partial x_j}+\frac{1}{a^2}\frac{\partial\rho}{\partial x_i}\frac{\partial U^\prime}{\partial x_j}]
\label{c9}
\end{eqnarray}
where $\Psi$ is the NC correction. Notice that the NC corrections usher in a form of inhomogeneity due to the non-trivial $x$-dependence.

 On the other hand, isolating the ${\bf{x}}$-dependent terms in (\ref{c7}) yield
\begin{eqnarray}
[\ddot{a}+\frac{4\pi}{3}aG\rho]x_i-\frac{1}{2a^3}\theta^{mj}\frac{\partial}{\partial x_i}(\frac{\partial\theta}{\partial x_j})\dot{a}^2 x_m =0,
\label{cc9}
\end{eqnarray}
 From (\ref{c9},\ref{cc9}) it is clear that  even the so-called "unperturbed" universe with ${\bf{u}}=0$ receives NC contributions. From (\ref{c9},\ref{cc9}) we recover
\begin{eqnarray}
\frac{1}{2}\frac{d}{dt}(\dot{a}^2)=\frac{4\pi G}{3}\frac{d}{dt}(\rho a^2) +\frac{4\pi G}{3}\Psi a^2 +\frac{1}{2a^3}\theta^{mj}\frac{\partial}{\partial x_i}(\frac{\partial\theta}{\partial x_j})\dot{a}^3 \frac{x_m x^i}{x^2},
\label{nc19}
\end{eqnarray}
leading to  a modified Friedmann equation,
\begin{eqnarray}
\frac{\dot{a}^2}{a^2} =\frac{8\pi G \rho}{3} -\frac{k}{a^2}+\frac{8\pi G }{3}\frac{1}{a^2}  \int a^2 \Psi dt+ \frac{\theta^{mj}}{a^2} \frac{x_m x^i}{x^2}\int\frac{\partial}{\partial x_i}(\frac{\partial\theta}{\partial x_j}) \frac{\dot{a}^3}{a^3} dt,
\label{c10}
\end{eqnarray}
where $k$ is a constant. In fact $k$ appears as an integration constant and can be identified with (the scaled) curvature in FRW. It will be more appealing to rewrite (\ref{c10}) as
\begin{eqnarray}
\frac{\dot{a}^2}{a^2} =\frac{8\pi G \rho}{3} -\frac{k_{eff}}{a^2},
\label{c100}
\end{eqnarray}
where 
\begin{eqnarray}
k_{eff} = k-\frac{8\pi G}{3} \theta^{ij}  \int \partial_i  \left[ \rho\dot{a}^2 x_j+\frac{1}{2a}\partial_k(\rho \dot{a} x_k)\partial_j \theta+\rho\partial_j U^\prime\right] dt - \theta^{mj} \frac{x_m x^i}{x^2}\int\partial_i\partial_j \theta \frac{\dot{a}^3}{a^3} dt.
\label{u5}
\end{eqnarray} 
The above is one of our principal results. We show that the NC-contribution can affect the flatness (or openness or closedness for that matter) although numerical estimates for the NC parameter $\theta^{ij}$ is needed to explicitly evaluate $k_{eff}$.

Again rewriting (\ref{cc9}) as below,	
\begin{eqnarray}
\frac{\ddot{a}}{a}=-\frac{4\pi G}{3}\rho+\frac{1}{2}\theta^{mj}\frac{\partial}{\partial x_i}(\frac{\partial\theta}{\partial x_j})\frac{x_m x^i}{x^2}\frac{\dot{a}^2}{a^4} 
\label{u1}
\end{eqnarray}
it is clear that the NC term acts as an effective pressure since 
 this equation summarizes the physics which determines the expansion of the universe. Hence even for vanishing conventional form of pressure the NC contribution can control the    acceleration or deceleration of the universe.
 
 The remaining  theoretical issue we pick up now is the following: how to define averages in Newtonian cosmology since to fit in the FRW framework we need to integrate out $x$ consistently. This has been developed by Buchert and coworkers in a series of papers \cite{buc,buc1,buc2} and is interpreted by them as an additional source in the form of   back reaction. This induced anisotropy and inhomogeneity can play important roles in structure formation. We will outline this  in the next section.	

 \section{Averaging Prescription in Cosmology}
A few crucial points have emerged from the works \cite{buc,buc1,buc2}: (i) the Friedmann equations are to be expressed in terms of spatially averaged variable such as averaged scale factor.\\
(ii) The averaged variables are required to be {\it{scalars}} to get unambiguous results in General Relativity since in general the procedure of averaging an inhomogeneous metric is not available. However for scalars spatial averaging can be properly defined for a foliated spacetime.\\
(iii) In standard Newtonian cosmology there is no non-vanishing {\it{global}}  averages since   the space is considered as boundaryless ($3$-torus) and the so called inhomogeneous back reaction terms reduce to surface contributions and hence vanish.\\
(iv) 	As emphasized by \cite{buc,buc1,buc4}, globally vanishing (averaged out) inhomogeneities  can contribute  as {\it{regional fluctuations}}. This is known as "cosmic variance". In this section we will be looking at NC-induced fluctuations.

As discussed earlier (\ref{a1}, see also \cite{das}) we will be working with the canonical variables by exploiting the Darboux map with no vorticity constraint  $v^i=\partial_i\theta$. From the  NC  fluid Lagrangian \cite{das}
\begin{equation}
L=-\partial_t\theta (\rho -\frac{1}{2}\theta^{ij}\partial_i\rho\partial_j\theta ) -(\frac{1}{2}\rho (\partial_i\theta)^2 +U(\rho) )
\label{x1}
\end{equation}
 the continuity and Euler equations follow:
\begin{eqnarray}
\partial_t\rho&=&\partial_l[-\rho(\partial_l\theta-\frac{1}{2}\theta^{lj}\partial_j\rho\frac{1}{\rho}(\frac{1}{2}(\partial_i\theta)^2+U^\prime)-\frac{1}{2}\theta^{ij}\frac{(\partial_j\rho)}{\rho}\partial_i\theta\partial_l\theta )]+O(\theta^3) \nonumber \\
&=&-\partial_l[\rho v^l]
\label{x4}
\end{eqnarray}
where,\begin{eqnarray} {v}^l=(\partial_l\theta-\frac{1}{2}\theta^{lj}\partial_j\rho\frac{1}{\rho}(\frac{1}{2}(\partial_i\theta)^2+U^\prime)-\frac{1}{2}\theta^{ij}\frac{(\partial_j\rho)}{\rho}\partial_i\theta\partial_l\theta ) +O(\theta^3) =v^l_c+O(\theta^{ij}),
\label{x5}
\end{eqnarray}
and
\begin{eqnarray}
\partial_t v^l=-\partial_l(\frac{{v}^2}{2})-\frac{1}{\rho}\partial_l P +\frac{1}{2\rho}\theta^{ij}\partial_l({v^i}\partial_jU )-\frac{1}{2}\theta^{ij}U^\prime\partial_l(\frac{1}{\rho}{v^i}\partial_j\rho )+\frac{1}{2}\theta^{lj}[U^\prime\partial_j(\frac{\partial_k(\rho {v^k})}{\rho})+\frac{{v^k}\partial_j\rho \partial_kU^\prime}{\rho}].
\label{x6}
\end{eqnarray}	 
with $P=\rho U^\prime-U$ as pressure.

It is straightforward to express the tensor $\partial_j v_i\lvert_c$ in terms of   rate of expansion $\psi=\nabla. v\lvert_c$, shear ($\sigma_{ij}$) are defined via the relation,
\begin{eqnarray}
\partial_j v_i\lvert_c=\sigma_{ij}+\frac{1}{3}\delta_{ij}\psi
\label{x7}
\end{eqnarray} where, $\sigma_{ij}=\frac{1}{2} (\partial_j v_i+\partial_i v_j)\lvert_c-\frac{1}{3}\delta_{ij} \psi$. Note that we have not taken in to account the anti-symmetric vorticity term. Rewriting  (\ref{x4}) and (\ref{x6}) in terms of the scalars $\rho,~\psi,~\sigma = \sqrt{\sigma^{ij}\sigma_{ij}}$ and using the convective time derivative operator $\dot A \equiv \frac{dA}{dt}= \partial_tA+\dot{{\bf{X}}} .\nabla A $ on a generic variable $A$,  we find,
\begin{eqnarray}
\dot{\rho}=-\rho\psi ,
\label{x8}
\end{eqnarray}
\begin{eqnarray}
\dot{\psi}&=&-\frac{1}{3}\psi^2-2\sigma^2 \nonumber \\
&&+\partial_l\left[-\frac{1}{\rho}\partial_l P +\frac{1}{2\rho}\theta^{ij}\partial_l({\partial_i\theta}\partial_jU )-\frac{1}{2}\theta^{ij}U^\prime\partial_l(\frac{1}{\rho}{\partial_i\theta}\partial_j\rho )+\frac{1}{2}\theta^{lj}[U^\prime\partial_j(\frac{\partial_k(\rho {\partial_k\theta})}{\rho})+\frac{{\partial_k\theta}\partial_j\rho \partial_kU^\prime}{\rho}]\right].
\label{x10}
\end{eqnarray}

Let us denote  $\langle A\rangle_D=\frac{1}{V}\int_D d^3x  A$ as the spatial average of a tensor field $A$ on the domain $D(t)$ occupied by the amount of fluid considered, and $a(t)$ is the scale factor of that domain. The subscript $D$ in $\langle A \rangle _D$ underlines the fact that it is not simply the  local $A$   that is averaged,
but a new domain dependent volume average of $A$. This is true for all subsequent definitions{\footnote{We thank Thomas Buchert for pointing this out to us.}}.

 We use the commutation rule for averaging \cite{buc1},
\begin{equation}
\langle A\dot{\rangle}_D-\langle \dot{ A}\rangle_D = \langle  A \psi\rangle_D -\langle A\rangle_D \langle \psi \rangle_D .
\label{com}
\end{equation} 

After averaging (using the commutation rule), the equations (\ref{x8}) and (\ref{x10}) becomes,
\begin{eqnarray}
\langle\rho\dot{\rangle}_D=-\langle\rho\rangle_D\langle\psi\rangle_D ,
\label{x11}
\end{eqnarray}
\begin{eqnarray}
\langle\psi\dot{\rangle}_D&=&\frac{2}{3}\langle\psi^2\rangle_D-\langle\psi\rangle^2_D-2\langle\sigma^2\rangle_D \nonumber \\
&&+\langle\partial_l\left[-\frac{1}{\rho}\partial_l P +\frac{1}{2\rho}\theta^{ij}\partial_l({\partial_i\theta}\partial_jU )-\frac{1}{2}\theta^{ij}U^\prime\partial_l(\frac{1}{\rho}{\partial_i\theta}\partial_j\rho )+\frac{1}{2}\theta^{lj}[U^\prime\partial_j(\frac{\partial_k(\rho {\partial_k\theta})}{\rho})+\frac{{\partial_k\theta}\partial_j\rho \partial_kU^\prime}{\rho}]\right]\rangle_D \nonumber \\
&=&\frac{2}{3}\langle\psi^2\rangle_D-\langle\psi\rangle^2_D-2\langle\sigma^2\rangle_D+\langle\partial_l(-\frac{1}{\rho}\partial_l P)\rangle_D \nonumber \\
&&+\langle\frac{1}{2}\epsilon_{ijr}\theta_r\left[\left(\frac{\partial_l\partial_j U}{\rho}-U^\prime\partial_l(\frac{\partial_j\rho}{\rho})\right)(\sigma_{il}+\frac{1}{3}\delta_{il}\psi)+\partial_i\left(U^\prime\frac{\partial_k\rho}{\rho}(\sigma_{kj}+\frac{1}{3}\delta_{kj}\psi)+U^\prime\partial_j\psi\right)\right]\rangle_D \nonumber \\
&& +\langle\frac{1}{2}\epsilon_{ijr}\theta_r\left[(\sigma_{ki}+\frac{1}{3}\delta_{ki}\psi)\left(U^\prime\partial_j(\frac{\partial_k\rho}{\rho})+\frac{\partial_j\rho\partial_kU^\prime}{\rho}\right)
\right]\rangle_D \nonumber \\
&& +\langle\frac{1}{2}\epsilon_{ijr}\theta_r\left[\partial_i\theta\partial_l\left(\frac{\partial_l\partial_j U}{\rho}-U^\prime\partial_l(\frac{\partial_j\rho}{\rho})\right)+\partial_l\theta\partial_i\left(U^\prime\partial_j(\frac{\partial_l\rho}{\rho})+\frac{\partial_j\rho\partial_lU^\prime}{\rho}\right)\right]\rangle_D ,
\label{x12}
\end{eqnarray}

 Now from (\ref{x12}), in terms of the volume scale factor $ a_D(t)$ (where $\langle\psi\rangle_D=3 \frac{\dot{ a_D}}{ a_D }$), the averaged Raychaudhuri equation can be written as,
 \begin{eqnarray}
 3\frac{\ddot{ a_D }}{ a_D}&=&\frac{2}{3}(\langle\psi^2\rangle_D-\langle\psi\rangle^2_D)-2\langle\sigma^2\rangle_D+\langle\partial_l(-\frac{1}{\rho}\partial_l P)\rangle_D \nonumber \\
 &&+\langle\frac{1}{2}\epsilon_{ijr}\theta_r\left[\left(\frac{\partial_l\partial_j U}{\rho}-U^\prime\partial_l(\frac{\partial_j\rho}{\rho})\right)\sigma_{il}+\partial_i\left(U^\prime\frac{\partial_k\rho}{\rho}\sigma_{kj}+U^\prime\partial_j\psi\right)\right]\rangle_D \nonumber \\
 && +\langle\frac{1}{2}\epsilon_{ijr}\theta_r\left[\sigma_{ki}\left(U^\prime\partial_j(\frac{\partial_k\rho}{\rho})+\frac{\partial_j\rho\partial_kU^\prime}{\rho}\right)
 \right]\rangle_D \nonumber \\
 && +\langle\frac{1}{2}\epsilon_{ijr}\theta_r\left[\partial_i\theta\partial_l\left(\frac{\partial_l\partial_j U}{\rho}-U^\prime\partial_l(\frac{\partial_j\rho}{\rho})\right)+\partial_l\theta\partial_i\left(U^\prime\partial_j(\frac{\partial_l\rho}{\rho})+\frac{\partial_j\rho\partial_lU^\prime}{\rho}\right)\right]\rangle_D
 \label{x14}
 \end{eqnarray}
For $\theta_i=0$ we recover earlier results of \cite{buc,buc1} and hence our NC model provides additional contributions to inhomogeneity and anisotropy.

As a specific example, let us consider a simple canonical dust form of barotropic potential   $U(\rho)=\nu \rho$ (with a constant $\nu$) that is pressureless,   $P=\rho U^\prime-U=0$. Further we impose no vorticity condition on velocity, ( a reasonable restriction valid at least until the epoch of structure
formation), leading $\psi=\partial_i(\partial_i\theta)=\partial^2\theta$ and $\sigma_{ij}=\partial_i\partial_j\theta-\frac{1}{3}\delta_{ij} \partial^2\theta$.
 The acceleration equation (\ref{x14}) reduces to,
\begin{eqnarray}
	3\frac{\ddot{ a_D }}{ a_D}&=&\frac{2}{3}(\langle\psi^2\rangle_D-\langle\psi\rangle^2_D)-2\langle\sigma^2\rangle_D \nonumber \\
	&&-\langle\frac{1}{2}\epsilon_{ijr}\theta_r\left[\partial_l(\frac{1}{\rho})\partial_j\rho\partial_i\partial_l\theta+\partial_i\theta\partial_l(\partial_j\rho\partial_l(\frac{1}{\rho}))+\frac{\partial_j\rho}{3\rho}\partial_i(\partial^2\theta)\right]\rangle_D .
	\label{p1}
\end{eqnarray}
In the next section we try to quantify the NC effect on cosmological parameters to bring out the implications of the NC extension.

 \section{Cosmological parameters with noncommutative corrections}
 It is clear from (\ref{x14}) that  inhomogeneities will act as  sources that will control  the average expansion rate of universe. Already such fluctuations have appeared in \cite{buc,buc1} and we have presented additional contributions generated by noncommutativity. After integrating (\ref{x14}), with $ k_D $ entering as an integration constant the resulting equation can be expressed in terms of cosmological parameters,
 \begin{eqnarray}
 \Omega_m^D+\Omega_k^D+\Omega_{NQ}^D=1.
 \label{p2}
 \end{eqnarray}
 where,
 $$\Omega_m^D=\frac{8\pi G\langle\rho\rangle_D}{3 H^2_D},~~ \Omega_k^D=-\frac{ k_D}{ a^2_D H^2_D},~\Omega_{Q}^D =\frac{2}{3}(\langle\psi^2\rangle_D-\langle\psi\rangle^2_D)-2\langle\sigma^2\rangle_D$$ and,
 \begin{eqnarray}
 \Omega_{NQ}^D=\Omega_{Q}^D +\frac{8\pi G}{3  a^2_D H^2_D}\theta^{ij}\langle\int \partial_i[\dot{a}^2\rho x_j +\rho\partial_jU^\prime+\frac{1}{2a}\partial_k(\rho\dot{a}x_k)\partial_j\theta] dt \rangle_D+\frac{\theta^{mj}}{ a^2_D H^2_D}\langle\int \left( (\frac{\dot{a}}{a})^3\frac{x_m x^i}{x^2}\partial_i\partial_j\theta\right)    dt\rangle_D .
 \label{p3}
 \end{eqnarray}
 Note that $\Omega_{Q}^D$ part was revealed in \cite{buc,buc1}.  However, there is an important distinction that all the parameters are domain dependent averages. In fact the integration constant or effective curvature $ k_D $ can also vary depending on the domain of averaging \cite{buc}. Note that in  (\ref{p3}) we have not considered the contribution coming from the Cosmological Constant. Apart from the conventional ones, $ \Omega_m^D,~\Omega_k^D $ there appears $\Omega_Q^D $, a form of kinematical back reaction effect. In standard FRW cosmology back reaction is absent and  a critical universe, $\Omega_m^D =1$ would have resulted in a flat universe with $ k_D =0$. Depending on the signature of the NC contribution, for positive sign, it can act as a kinematical dark matter thereby enhancing structure formation. On the other hand a negative contribution can play the role of kinematical Cosmological Constant that  will favor accelerated expansion.

 Another way of  writing the Friedmann equation involves  the present day expansion and density parameters. At $t=t_0$, $a_D= a_{0D}$ the Friedmann equation (\ref{c10}) takes the form
 \begin{eqnarray}
 \frac{ k_D}{ a_{0D}^2}&=&\frac{8\pi G\langle\rho_0\rangle_D}{3}- H_{0D}^2+\frac{8\pi G}{3  a_{0D}^2}\theta^{ij}\langle \int \partial_i[\rho(t)x_j \dot{a}^2+\rho(t)\partial_jU^\prime +\frac{1}{2a}\partial_k(\rho(t)\dot{a}x_k)\partial_j\theta] dt\lvert_{t=t_0} \rangle_D \nonumber \\
 &&+\frac{\theta^{mj}}{ a_{0D}^2}\langle\int \left((\frac{\dot{a}}{a})^3 \frac{x_m x^i}{x^2}\partial_i\partial_j\theta\right)    dt\lvert_{t=t_0}\rangle_D  \\
 &=& H_{0D}^2\left[\Omega_{m0}^D-1+\frac{8\pi G}{3  a_{0D}^2 H_{0D}^2}\theta^{ij}\langle\int \partial_i[\dot{a}^2\rho(t)x_j +\rho(t)\partial_jU^\prime+\frac{1}{2a}\partial_k(\rho(t)\dot{a}x_k)\partial_j\theta] dt\lvert_{t=t_0} \right]\rangle_D \nonumber \\
 &&+H_{0D}^2\left[\frac{\theta^{mj}}{ a_{0D}^2 H_{0D}^2}\langle \int \left( (\frac{\dot{a}}{a})^3\frac{x_m x^i}{x^2}\partial_i\partial_j\theta\right)    dt\lvert_{t=t_0}\rangle_D \right]
 \label{u2}
 \end{eqnarray}
 where $\Omega_{m0}^D=\frac{8\pi G\langle\rho_0\rangle_D}{3 H_{0D}^2}$ and $ H_{0D}= \frac{\dot{a}_D}{a_D}\lvert_{t=t_0}$.
 
 The curvature parameter $k_D$ can be eliminated from the Friedmann equation thereby yielding
 \begin{eqnarray}
 \left(\frac{\dot{a}_D}{ a_D}\right)^2
 &=& H_{0D}^2\left[\frac{8\pi G}{3H_{0D}^2}\left(\frac{a_{0D}^3}{a_D^3}\langle\rho_0\rangle_D+\frac{\theta^{ij}}{ a^2_D}\langle\int \partial_i[\dot{a}^2\rho x_j +\rho\partial_jU^\prime+\frac{1}{2a}\partial_k(\rho\dot{a}x_k)\partial_j\theta] dt\rangle_D\right)\right] \nonumber \\
 &&+H_{0D}^2\left[\frac{\theta^{mj}}{ a^2_D H_{0D}^2}\langle\int \left( (\frac{\dot{a}}{a})^3\frac{x_m x^i}{x^2}\partial_i\partial_j\theta\right)    dt\rangle_D\right] \nonumber \\ &&+H_{0D}^2\left[\frac{ a_{0D}^2}{ a^2_D}\left(1-\Omega_{NC}^D\right)\right]
 \label{u3}
 \end{eqnarray}
 where,
 \begin{eqnarray}
 \Omega_{NC}^D&=&\Omega_{m0}^D+\frac{8\pi G}{3  a_{0D}^2 H_{0D}^2}\theta^{ij}\langle\int \partial_i[\dot{a}^2\rho(t)x_j +\rho(t)\partial_jU^\prime+\frac{1}{2a}\partial_k(\rho(t)\dot{a}x_k)\partial_j\theta] dt \lvert_{t=t_0}\rangle_D \nonumber \\
 &&+\frac{\theta^{mj}}{ a_{0D}^2 H_{0D}^2}\langle\int \left( (\frac{\dot{a}}{a})^3\frac{x_m x^i}{x^2}\partial_i\partial_j\theta\right)    dt\lvert_{t=t_0}\rangle_D .
 \label{u4}
 \end{eqnarray}
 
 An interesting point to note is that in (\ref{u3}) the NC contribution affects both Friedmann equation as well as the constant parameter $ \Omega_{NC}^D$.  This constitutes another of our major result where we explicitly provide contribution of the NC correction term in the energy budget of the universe.

To further quantify the inhomogeneity effects one needs to assume specific forms of the velocity such as in spherical collapse model (with a spherically symmetric and radial velocity inside the averaging domain) or in Eulerian linear approximation (where Eulerian coordinates are comoving with the background Hubble flow) \cite{buc}. We plan to explore these aspects in a separate publication. 	
\section{Discussion and summary}

In this perspective let us finally discuss possible observational consequences of the results presented in our model. To compare and contrast with the above mentioned scenario we note that the tensorial constant NC parameter $\theta_{ij}$ can be identified with the two-form field introduced above but instead of exploiting the locally rotationally symmetric class of metrics as above we have considered NC-extended FRW metric. As we have already advertised in the Introduction, there are essentially two aspects where noncommutativity in the fluid can have effects as seen in (\ref{u2}) (NC-corrected effective curvature) and (\ref{u4}) (NC-corrected effective energy budget). As we have already noted  (\ref{u4}) indicates that NC-effect can contribute additionally as either  Dark Matter or Dark Energy,  which, however  is more of theoretical interest. On the other hand (\ref{u2}) shows that noncommutativity can have a direct impact on observational cosmology and hence we will slightly elaborate on this issue of topical interest. Quite obviously NC effect induces {\it {anisotropy}} since the NC parameter $\theta_{ij}=\epsilon_{ijk}\theta_k$ being a constant tensor introduces a preferred spatial direction. As we have discussed briefly, our model provides a form of 
anisotropy that is  quite subtle since it is (in some sense) concealed   within an effective FRW framework because the background  expands isotropically and CMB is also isotropic at the background level. 

There is quite a fair amount of work in the context of NC effects in Cosmology. We have listed some of them in \cite{nccosmo,bala1,nccmb}. It needs to be mentioned that different NC structures are exploited in these works ranging from the string theory motivated constant NC, (parametrized by a constant $\theta_{ij}$, considered by us as well), to the more complicated operatorial form of non-constant NC structures (stemming from Generalized Uncertainty Principle framework). However there is consensus among different NC practitioners that NC effects should be {\it{directly}} relevant at the ultraviolet, i.e.  extremely  high energy or short distance (such as Planck energy or length scale maybe) but interestingly its {\it{indirect}} signature can be present at the infrared, i.e.  in low energy or long wavelength regime. The prevalent idea is that close to the Big Bang era, the energy density or dimension of the universe were (presumably) around Planck scale so that NC could have impacted directly but its indirect impact can be felt at present time in low energy cosmology in the form of anisotropies in CMB or (as pointed out here) in direction dependent  effective curvature or energy content. In is worthwhile to emphasize that introduction of NC effects from noncommutative fluid perspective  that has been done in the present work is completely since the NC extension of fluid has been a recent development, initiated by us \cite{pra,das,ar1}.

The existing bound on the NC parameter is $\mid\theta_{ij}\mid\approx  \leq(10 GEV)^{-2}  $ \cite{theta}.
In \cite{bala1} Balachandran et.al. have studied quantum fluctuations of the inflaton scalar field on certain NC spacetimes and NC effects in anisotropies in  CMB radiation,   large scale structure (of matter also), as well as NC-induced direction-dependence in  power spectrum. In \cite{nccmb} the above authors have attempted to constrain the (theoretical) NC length scale to  around $10~ TeV$ by matching with the
observational data from ACBAR, CBI and 5 year WMAP that tentatively fixes the  
scale factor $a(t)$ at the end of  inflation. It is interesting to note that we have also derived similar NC effects in cosmological observables. These values can be used in (\ref{a1}) for specific models to come up with numerical estimates of NC effects although it is expected that the NC corrections might be very small.

Let us conclude by summarizing the present work with a mention of possible future directions. In the first part of the paper we have generalized our previously proposed noncommutative fluid model by including fluid vorticity. Detailed discussions on the extended Hamiltonian structure (Dirac brackets) and dynamics have been provided. Major part of the paper deals with cosmological implications of the noncommutative extension, especially in terms of generating inhomogeneity and anisotropy. Following the formalism developed by Buchert and coworkers \cite{buc,buc1} we show that the nocommutative effects induce additional back reaction terms that can affect cosmological evolution.

As remaining open problems we plan to estimate the noncommutative back reaction effect quantitatively by utilizing specific models of fluid motion. A more ambitious project is to extend the noncommutative fluid model in relativistic scenario. So far our noncommutative extension is essentially non-relativistic that has forced us to consider Newtonian  cosmology. Indeed a fully relativistic noncommutative fluid model will open up many new questions and challenges.

\vskip .5cm
{\bf {Appendix}}:\\
 In a generic Second Class Constraint system with $n$  SCCs $\chi_i$, $i=1,2,..n$, the modified symplectic structure (or Dirac brackets) are defined in the following way,
 \begin{equation}
 \{A,B\}^*=\{A,B\}-\{A,\chi _i\}\{\chi ^i,\chi ^j\}^{-1}\{\chi _j,B\}, \label{a3}
 \end{equation}
 where $\{\chi ^i,\chi ^j\}$ is the invertible constraint matrix.  Denoting the canonically conjugate momentum of a generic variable $a$  by $\Pi_a$, the Second Class Constraints  $\chi_i$, $i=1,2,3,4$ are be computed from the Lagrangian:
\begin{eqnarray}
\chi_1= \Pi_\theta+\rho-\frac{1}{2}\theta^{ij}\partial_i\rho\partial_j\theta \nonumber \\
\chi_2=\Pi_\alpha \nonumber \\
\chi_3=\Pi_\beta+\rho \alpha \nonumber \\
\chi_4=\Pi_\rho 
\label{c1}
\end{eqnarray}
The constraint matrix $\{\chi_i,\chi_j\}$ ( where $i$ and $j$ goes from 1 to 4) can be written as,
\begin{eqnarray} \{\chi_i,\chi_j\}=
\begin{pmatrix}
-\theta^{ij} \partial_i\rho\partial_j\delta (x-y) & 0 & 0 & \delta (x-y)-\frac{1}{2} \theta^{ij} \partial_j\theta\partial_i\delta (x-y)  \\
0 & 0 & -\rho\delta (x-y) &  0 \\
0 & \rho\delta (x-y)  &  0 & \alpha\delta (x-y) \\
-\delta -\frac{1}{2} \theta^{ij} \partial_j\theta\partial_i\delta (x-y) & 0 &  -\alpha\delta (x-y) & 0 
\end{pmatrix}
\label{c2}
\end{eqnarray}
where all field arguments and derivatives on $\delta (x-y)$ are on $x$. Due to the Second Class nature of the constraints, the constraint matrix $\{\chi_i,\chi_j\}$ is  invertible and the inverse, to $O(\theta^{ij})$,  reads as,

\begin{eqnarray}\{\chi ^i,\chi ^j\}^{-1}=
\begin{pmatrix}
	 0 & C(x,y)  & 0 & D(x,y)  \\
	-C (x\rightleftharpoons y)  & - \theta^{ij}\frac{\alpha(x) \alpha(y)}{\rho(x)\rho(y)}\partial_i\rho(x)\partial_j\delta(x-y) & \frac{\delta(x-y)}{\rho} & E(x,y)  \\
	0 & -\frac{\delta(x-y)}{\rho} & 0 & 0 \\
	-D (x \rightleftharpoons y) & -E (x \rightleftharpoons y)  & 0 & -\theta^{ij} \partial_i\rho(x) \partial_j\delta(x-y)   
	\end{pmatrix}
\end{eqnarray}
where,

 $C=\frac{\alpha}{\rho} \delta(x-y)-\frac{1}{2} \theta^{ij} \frac{\alpha (y)}{\rho (y)} \partial_j\theta (x)\partial_i\delta (x-y)$,  $D=-\delta(x-y)+\frac{1}{2} \theta^{ij}  \partial_j\theta (x)\partial_i\delta (x-y)$, \\
  $E=\theta^{ij} \frac{\alpha}{\rho} \partial_i\rho(x) \partial_j\delta(x-y)$  and $(x \leftrightharpoons y)$ means the argument at $x$ goes to $y$. \\

The Dirac brackets follow from using (\ref{a3}):
\begin{eqnarray}
\{\rho(x),v^i(y)\}=-\partial_i\delta (x-y)-\frac{1}{2}\theta^{kl}\partial_l\theta(x)\partial_i\partial_k\delta(x-y)+\theta^{kl}\partial_k\rho[\partial_l(\frac{\alpha}{\rho}\partial_i\beta)\delta (x-y)+\frac{\alpha}{\rho}\partial_i\beta\partial_l\delta(x-y)],
\label{nc3} 
\end{eqnarray}	
\begin{eqnarray}
\{v^i(x),v^j(y)\}&=&\frac{\delta(x-y)}{\rho} [\partial_j v^i-\partial_i v^j] \nonumber \\
&&
 +
 \frac{1}{2} \theta^{kl}\left[\frac{\alpha}{\rho}\partial_l\theta\partial_i\beta \partial_j\partial_k\delta(x-y)-\partial_i\left(\partial_l\theta(\partial_k(\frac{\alpha}{\rho}\partial_j\beta)\delta(x-y)+\frac{\alpha}{\rho}\partial_j\beta\partial_k\delta(x-y))\right)\right] \nonumber \\
&&
 -\theta^{kl}\left[\frac{\alpha}{\rho}\partial_i\beta\partial_k\rho\left(\partial_j\beta\partial_l(\frac{\alpha}{\rho})\delta(x-y)+\frac{\alpha}{\rho}\partial_j\beta\partial_l\delta(x-y)+\frac{\alpha}{\rho}\partial_l\partial_j\beta \delta(x-y)\right)\right]
\label{nc5}
\end{eqnarray}

\vskip .5cm
\textbf{Acknowledgement}: It is a pleasure to thank Thomas Buchert for his helpful comments and for sending us reprints of early literature. We are grateful to the referees for constructive comments. The work of  P. D. is supported by INSPIRE, DST, India.

	\end{document}